\documentclass[11pt]{article}
\usepackage{graphicx}
\def\m{\mu}

\def\l{\lambda}

\def\ra{\rangle}
\def\la{\langle}

\def\ZG{{\cal Z}_G(L,s)}

\begin{document}

\title{A new class of models for surface relaxation\\ with\\ exact mean-field solutions}
\author{V. Karimipour \footnote{e-mail: vahid@sina.sharif.ac.ir.}
\\ \\
B. H. Seradjeh  \footnote{e-mail:
babak@sfu.ca}\\ \\
Department of Physics, Sharif University of Technology,\\
                         P. O. Box 11365-9161, Tehran, Iran \\}

\maketitle
\begin{abstract}
We introduce a class of discrete models for surface relaxation.
By exactly solving the master equation which governs the
microscopic dynamics of the surface, we determine the steady state
of the surface  and calculate its roughness. We will also map our
model to a diffusive system of particles on a ring and
reinterpret our results in this new setting.

\end {abstract}
\textbf{Keywords:} Stochastic process, matrix product ansatz,
zero range contact process, surface relaxation.

\section{Introduction}
 Models far from thermodynamic equilibrium, may not
obey detailed balance and hence their steady state can not be
described by Gibbs Measure, and one often needs to study closely
the stochastic dynamics and obtain the steady state directly by
simulation, numerical or exact solution of the microscopic
dynamical equations.(For recent reviews see \cite{hi,sr} and
references therein). The microscopic dynamics which is usually
encoded in a set of transition rates between different
configurations, usually bring about strong long-range
correlations in one dimensional systems. This is the case, for
example in driven diffusive systems the prototype of which is the
Asymmetric Simple Exclusion Process\cite{sp}. In such cases then,
the mean field solution is not exact or else is exact only under
certain conditions on the rates.

In recent years, much attention has been paid to the exact
solution of stochastic processes on one dimensional lattices.
Among the methods developed, the Matrix Product Ansatz (MPA), has
proved particularly fruitful, since it has both led to simple and
nice solutions of the previously known models \cite{de} and has
also led to various generalizations [5-12].
%\cite{ahr,ev,evg,li,k1,k2,kk,deg}.

The Matrix Product Ansatz is best understood within the operator
formalism for Markov processes on one dimensional chains in which
the steady state of the process, turns out to be the ground state
of a suitable Hamiltonian with nearest neighbor interactions
\begin{equation}\label{1}
  H = \sum_{k=1}^L h_{k,k+1}.
\end{equation}
According to MPA, the steady state probability $ P(\tau_1,
\tau_2, \cdots \tau_L)$ for the configuration $ (\tau_1, \tau_2,
\cdots \tau_L)$ is written as the trace of a product of
noncommuting matrices:
\begin{equation}\label{2}
  P(\tau_1, \tau_2, \cdots \tau_L)= \frac{1}{Z(L)}tr\Bigg(D(\tau_1)
  D(\tau_2)\cdots D(\tau_L)\Bigg)
\end{equation}
(Here we are taking a closed system with periodic boundary
conditions.) It is then easy to show that the above probabilities
are the steady state ones, provided that the matrices $D(\tau)$
satisfy a set of relations encoded compactly as:\cite{ks}
\begin{equation}\label{3}
  h \textsf{D}\otimes\textsf{D}= \textsf{X}\otimes \textsf{D}- \textsf{D}
\otimes \textsf{X}\end{equation} Here $\textsf{D}$ and $
\textsf{X}$ are matrix valued column vectors.

The solution of these equations are however far from trivial and
so far relatively few systems have lend themselves to this type of
solution. In cases where we take $D(\tau)$ to be one dimensional
and accordingly denote them by $d({\tau})$, we are in fact
searching for exact mean field solutions. Even in this case,
(\ref{3}) is a set of  coupled nonlinear algebraic equations and
are generally difficult to solve, or else so many constraints of
consistency are imposed on the rates to render the process
uninteresting and
non-physical.\\
One can however reverse the problem \cite{adr,k1,k2,r2} and search
among quadratic algebras for those classes that may correspond
through
the MPA formalism to a stochastic Hamiltonian.\\
Mathematically, a through search and classification is quite
difficult, and one is better to start from a restricted class of
algebras or processes which seem to lead to physically
interesting models.\\ ( For a partial classification of algebras,
which pertain to diffusion and exchange of particles see
\cite{r2}).\\
In this paper we take all the matrices to be one dimensional
which means we are considering those processes for which mean
field solutions are exact. As far as the number of particles of
each species on a ring is conserved, a one dimensional
representation leads to a rather uninteresting steady state,
namely, one in which all the configurations have equal weights.
However if there is no such conservation law, then these mean
field solution may be quite nontrivial and contain interesting
physics, as we shall see.

Our starting point is to write a set of microscopic dynamical
rules pertaining to a generalization of the so-called zero-range
contact process \cite{sp}(see also \cite{evr} for more references)
and ask if the steady state of the corresponding stochastic
process allows an exact mean field solution (i.e: a one
dimensional MPA solution). We find that an interesting and rather
general class of these processes allow such solutions, namely
those processes in which the rates of transitions between
neighboring sites can be factorized in two pieces each pertaining
to one of the sites . These processes can have two
interpretations, one in terms of surface relaxation and the other
in terms of diffusive systems of particles. For definiteness we
adhere to the first interpretation and determine exactly the
steady state of the surface and calculate the roughness of the
surface. We study in detail two members of the class.

\section{Models for surface relaxation}
We consider a ring of $L$ sites, each site of which can
accommodate a number of particles on top of each other. At most
$p$ particles can be accommodated in each site. To each site $k$
of the lattice ($k=1,\cdots L$), we assign a stochastic variable
$\tau_k$ which shows the number of particles in that site, i.e.
($\tau_k = 0,1,\cdots p$).

The microscopic stochastic dynamics which governs the relaxation
of the surface are described according to the following
transition rules between the stochastic variables pertaining to
any two adjacent sites $k, $ and $k+1$(Fig. 1):

\begin{eqnarray}\label{4,5}
i\quad j &\longrightarrow& i-1\quad j+1 \qquad\hbox{with rate}\ \
\ \ \m_{i,j},\\ j\quad i &\longrightarrow& j+1\quad i-1
\qquad\hbox{with rate}\ \ \ \ \m_{i,j},
\end{eqnarray}
where $i$ and $j$ represent the values of $ \tau_k, $ and
$\tau_{k+1}$.  Note that the model has left-right symmetry and we
are taking time to be continuous, hence we speak of rates rather
than probabilities.  Obviously we should have the following
constraints on the rates, namely $\mu_{0j} = \mu_{jp} = 0 \ \ \ \
\forall j $. Moreover we require that the higher the difference
between the heights of adjacent columns, the more probable is the
above transition. Thus we require that $\mu_{ij}$ be an
increasing function of $(i-j)$.\\
Note that the surface does not grow on the average and the total
number of particles is constant which we take to be equal to $N$.
If the relaxation of the surface is much faster than the rate of
fall of particles , then this model can also describe the growth
of the surface, in this case time will be proportional to the
number of particles $N$.

\section{Interpretation as a diffusive system of particles}
Our model can also be interpreted as a model for particles
hopping on a ring. One can now think of $L$ particles on a ring
of $L+N$ sites, where each particle $k$ has a number of vacant
sites $ \tau_k$ ahead. In this view, the transition rules
(\ref{4,5}) are interpreted as follows: A particle whose
distances to its left and (right) neighboring particles are $i$
and ($j$), hops forward with rate $\mu_{ji} $ and backward with
rate $\mu_{ij}$

The increasing property of $\mu_{i,j}$ in terms of $(i-j)$ is
still plausible in this setting, specially if we think of traffic
flow, since a car in the above condition accelerates if $i-j$ is
small and decelerates if $i-j$ is large. In this interpretation
the probabilites (2) give the distribution of headways and the
roughness designates how uniform the flow is. For reviews on
traffic flow and (particularly its statistical mechanical
aspects), see \cite{helb}, and (\cite{chad}) respectively.\\

The above process can now be studied as an interacting stochastic
system. For this study we use the operator formalism of Markov
processes, where to each configuration $ (\tau_1, \tau_2, \cdots
\tau_L$) of the system a state vector $|\tau_1, \tau_2, \cdots
\tau_L> $ is assigned and each probability distribution $P$, is
encoded in a vector
\begin{equation}\label{6}
|P> := \sum_{\tau_1,\tau_2,\cdots \tau_L}P(\tau_1,\tau_2,\cdots
\tau_L)|\tau_1,\tau_2,\cdots \tau_L)>.
\end{equation}
The totality of possible state vectors span a complex Hilbert
space, denoted by ${\cal H}:=(C^{p+1})^{\otimes L}$, where
$C^{p+1}$ is the complex space spanned by ${\cal
B}=\{\,|0\rangle,\,|1\ra\,\ldots\,|p\ra\}.$  The stochastic
Hamiltonian describing the above dynamical rules, is given by

\begin{equation}\label{7}
 {\cal H}=\sum_{k=1}^L h_{k,k+1},
\end{equation}
where
\begin{equation}\label{8}
 h=\sum_{i,j}
-\m_{i,j}\Bigg(|i-1\ j+1\ra\la i\ j| + |j+1\ i-1\ra\la j\ i| -|i\
j\ra \la i\ j|-|j\ i\ra \la j\ i|\Bigg).
\end{equation}

Inserting (\ref{8}) with the choice ${\textsf D} = column (d_0,
d_1, \cdots  d_p)$ and $ {\textsf X = 0} $ into equation (\ref{3})
we obtain :

\begin{equation}\label{9}
\m_{i+1,j-1}d_{i+1}d_{j-1}+\m_{j+1,i-1}d_{i-1}d_{j+1} =
(\m_{i,j}+\m_{j,i})d_id_j \qquad \forall i,j.
\end{equation}
A sufficient condition for solving these equation is the
following:
\begin{equation}\label{10}
\m_{j+1,i-1}d_{i-1}d_{j+1}= \m_{i,j}d_id_j \qquad \forall i,j,
\end{equation}

For general rates these relations may not have any non-zero
solutions, however if we take the rates to be of factorized form,
namely
\begin{equation}\label{11}
\m_{i,j} = \mu_i \lambda_j,
\end{equation}
where $\mu_i$ and $\lambda_j$ are arbitrary non-negative real
numbers, then the set of equations(\ref{10}) have the following
solution for $d_i$'s:
\begin{equation}\label{12}
  d_0 = 1,\ \ \ \ d_1 = \frac{\l_0}{\m_1}, \ \ \ \ d_2 =
  \frac{\l_1\l_0}{\m_2\m_1}\ \ \ \
  d_3 =  \frac{\l_2\l_1\l_0}{\m_3\m_2\m_1}\ \ \ \cdots\ \ \ \
  d_p = \frac{\l_{p-1} \cdots \l_1\l_0}{\m_p\cdots \m_2\m_1}
 \end{equation}
 \textbf{Remark:}\\
 The symmetry of the process is crucial in the
 above derivation, for a totally asymmetric process, the rates
 will be highly constrained.\\

 Note that the factorization property is the only requirement that
 exact solvability imposes on this model. We are free to choose the dependence of
$\mu_i$ and $\lambda_i$ on $i$ as we wish. It is for physical
reasons that we require that
\begin{eqnarray}\label{13,14}
\mu_p \geq \mu_{p-1}\ \cdots \mu_1 \geq \mu_0 &=& 0\\
\lambda_0 \geq \lambda_1 \geq \cdots \lambda_{p-1} \geq \lambda_p
&=& 0
\end{eqnarray}
In this way the rates $\mu_{i,j}$ will be increasing functions of
$ (i-j)$. Furthermore the probability of a particle jumping from
a lower height to a neighboring higher position will be quite low.
We can now calculate the probability of all configurations. From
(\ref{2}) we find:
\begin{equation}\label{15}
  P(\tau_1, \tau_2, \cdots \tau_L) = \frac{1}{{\cal Z}(L,N)}
  d(\tau_1) d(\tau_2) \cdots d(\tau_L)
\end{equation}
Where ${\cal Z}(L,N)$, the partition function, is given by:

\begin{equation}\label{16}
{\cal Z}(L,N):= \sum_{\sum_{k=1}^L \tau_k = N} d(\tau_1)
d(\tau_2) \cdots d(\tau_L)
\end{equation}
Due to the commutativity of $d(\tau)$'s, the probabilities depend
only on number of columns $K_i$ of a given height $i$, and not on
the position of columns. Thus (\ref{15}) can be rewritten as:
\begin{equation}\label{17}
  P(K_0, K_1,\cdots, K_p) = \frac{1}{{\cal Z}(L,N)}
\frac{L!}{K_0!K_1!\cdots K_p!}
  d_0^{K_0}
d_1^{K_1}\cdots d_p^{K_p},
\end{equation}
where the combinatorial term account for various possible places
of the columns, and (\ref{16}) can be re-expressed as:
\begin{equation}\label{18}
{\cal Z}(L,N):= \sum"\frac{L!}{K_0!K_1!\cdots K_p!} d_0^{K_0}
d_1^{K_1}\cdots d_p^{K_p}
\end{equation}
Where by $\sum" $ we mean a sum subject to $  \sum_{i=0}^p
iK_i=N,$ \ \ \ {\rm{and}} \ \ \ $ \sum_{i=0}^{p}K_i = L $. Our
analysis will be simplified if we consider a grand partition
function $ \ZG $:
\begin{eqnarray}\label{19,20,21}
\ZG &=& \sum_{N=0}^{pL} s^N {\cal Z}(L,N) \\
    &=& \sum_{N=0}^{pL}s^N \sum"
    \frac{L!}{K_0!K_1!\cdots K_p!}
     d_0^{K_0}d_1^{K_1}\ldots d_p^{K_p}\\
         &=& \left(d_0+d_1s+\ldots d_ps^p\right)^L =: (\textbf{z}(s))^L.
\end{eqnarray}
The parameter $s$ plays the role of fugacity. The average density
of particles ${n}:= \frac{N}{L}$ and the average ratio of columns
of height $i$, that is ; $\overline{k_i}:=
\frac{\overline{K_i}}{L}$, are given respectively by:
\begin{equation}\label{22}
n = \frac{s}{L}\frac{\partial}{\partial s}\ln\ZG =
s\frac{d}{ds}\ln {\bf{z}}(s)\hskip 2cm \overline{k_i} =
\frac{d_i}{L}\frac{\partial}{\partial d_i}\ln \ZG = \frac{d_i
s^i}{\textbf{z}(s)}
\end{equation}
The roughness of a surface is usually defined as \cite{st}:
\begin{equation}\label{23}
  \overline{W}:=\sqrt{\frac{1}{L}\sum_{x}\la(h(x)-\overline{h})^2\ra}
\end{equation}
where $h(x)$ is the height of the surface at point $x$. This can
be rewritten as:
\begin{equation}\label{24}
 \overline{W} =\sqrt{\frac{1}{L}\sum_{0}^p \overline{K_i}
  (i - \frac{\overline{N}}{L})^2}
\end{equation}
A simple calculation shows that:
\begin{equation}\label{25}
\overline{W}^2 = (s\frac{d}{ds})^2 \ln {\bf z}(s)=
(s\frac{d}{ds}){n}
\end{equation}

We now consider two examples which lend themselves to elementary
analysis.
\subsection{Example A}
In this first example, we take the rates to be as follows:
\begin{equation}\label{26}
  \mu_i = i \hskip 2cm \lambda_i = p-i
\end{equation}
The above assignments satisfy all the requirements put forward on
physical grounds. For example the rate of fall of a particle from
height $p$ to height $0$ is $p^2$, while the rate of jump of a
particle from a height $1$ to $p-1$ is $1$. Moreover $\mu_0
=\lambda_p = 0$. In this case we find from (\ref{12}):
\begin{equation}\label{27}
  d_i = \frac{p!}{i!(p-i)!}
\end{equation}
and
\begin{equation}\label{28}
  {\bf z}(s) = \sum_{i=0}^{p} \frac{p!}{i!(p-i)!}s^i = (1+s)^p
\end{equation}
From (\ref{22}) we find the average number of particles:
\begin{equation}\label{29}
{n}= \frac{ps}{1+s}\hskip 2cm {\rm or} \hskip 2cm s =
\frac{n}{p-n}
\end{equation}
Also from (\ref{22}) we find the fraction of sites with height $i$
to be:
\begin{equation}\label{30}
\overline{k_i}= \Bigg(\begin{array}{c}
  p \\  i\end{array} \Bigg) \frac{s^i}{(1+s)^p}
\end{equation}

Rewriting this in terms of the number of particles ${n}$, we find:
\begin{equation}\label{31}
\overline{k_i}=\frac{1}{p^p}\Bigg(\begin{array}{c}
  p \\  i\end{array} \Bigg)
{n^i}(p-n)^{p-i},
\end{equation}
which implies a binomial distribution for the columns of
different heights.\\
The roughness is calculated from (\ref{25}) and (\ref{28}) to be:
\begin{equation}\label{32}
  \overline{W} =\sqrt{{n}(1-\frac {{n}}{p})},
\end{equation}
If the relaxation is much faster than the rate of fall of
particles, we can take $n \sim t $ and $ p\sim T$ where $T$  is
the time for growing a layer of uniform thickness $p$, then the
above equation implies, that at the early stages of the growth
when $n\ll p$, the roughness increases as $ \approx
t^{\frac{1}{2}}$, in the middle stages of the growth, $n$ is
comparable with $p$, it increases linearly with time, and finally
at the final stages when the number of holes $p-n$ becomes a good
measure of time $T-t$, the roughness again
decreases as the square root of $T-t$.\\

\subsection{Example B:} In this example we consider the following
rates:
\begin{equation}\label{33}
\mu_i = 1 \hskip 0.5cm \forall i\ne 0, \hskip 2cm \lambda_i = 1,
\hskip 0.5cm \forall i\ne p \hskip 2cm \mu_0 = \lambda_p = 0
\end{equation}
That is, the rates do not depend on the height of source and
target columns. We find from (\ref{12}) that $d_0=d_1=\cdots d_p
= 1 $ and thus:
\begin{equation}\label{34}
 {\bf z}(s) = \frac{1-s^{p+1}}{1-s}
\end{equation}
from (\ref{22}) we find:
\begin{equation}\label{36}
\overline{k_i} = \frac{s^i(1-s)}{1-s^{p+1}}
\end{equation}
\textbf{Large p-limit} Some insight is obtained if we look at the
large
$p$ limit of the roughness  for the two cases $s<1$ and $s>1$.\\
When $s<1$, we find from (\ref{34}), that ${\bf z}(s)\simeq
\frac{1}{1-s}$ and
 using (\ref{22}) and (\ref{25}) we obtain:
\begin{equation}
{n}\simeq \frac{s}{1-s}\hskip 2cm  \overline{W}\simeq
\frac{\sqrt{s}}{(1-s)} = \sqrt{n(1+n)}
\end{equation}

On the  other hand when $ s>1$, we find that $\ln {\bf
z}(s)\simeq (p+1)\ln s - \ln(s-1)$, which when inserted into
(\ref{22}) and (\ref{25}), yields respectively
\begin{equation}\label{a}
n \simeq p - \frac{1}{s-1}\hskip 1cm {\rm {and}}\hskip 1cm
\overline{W}\simeq \frac{\sqrt{s}}{s-1}= \sqrt{(p-n)(p-n+1)}
\end{equation}
Moreover from (\ref{36}) we find that in the large $p$ limit,
when $s<1$  or  $n < p/2$,
\begin{equation}\label{c}
 \overline{k_i}\simeq \frac{n^i}{(1+n)^{i+1}}
\end{equation} and when $ s > 1$ or $n > p/2$
\begin{equation}
\overline{k_i}\simeq
    \frac{(p-n)^{p-i}}{(1+p-n)^{p-i+1}}
 \end{equation}
It is interesting to note that despite the very different nature
of the transition rates in models ${\bf A}$ and ${\bf B}$, in
both models, the formulas obtained for the roughness imply that
the roughness first increases as $t^{1/2}$ and then increases
linearly and finally decreases at the final stage again as
$(T-t)^{1/2} $.\\

These two  regimes are separated by the case when $s=1$,\  or\
${n}= \frac{p}{2}$, where we find for all arbitrary $p$'s,
$\overline{k_i} = \frac{1}{p+1}$, meaning that all heights are
equally probable. In this case the roughness turns out from
(\ref{25}) to be:
\begin{equation}\label{37}
\overline{W} = \sqrt{\frac{p^2}{12}+ \frac{p}{6}}.
\end{equation}

Fig.(3) shows the value of roughness in terms of the fugacity for
$p=10$. It is seen that there is a sharp peak in the roughness at
${n}= \frac{p}{2}$, when half of the capacity of the lattice is
filled
with particles.\\
All these hint that the point $ s=1$ or ${n}= \frac{p}{2}$ is a
special point, possibly a point of phase transition.\\ \\
{\Large\textbf{Acknowledgement:}} V. K. would like to thank V.
Rittenberg for his valuable comments. We would like to thank A.
Rezakhani for his valuable help in preparing the manuscript.
\newpage
{\large {\bf References}}

\newpage
\begin{figure}
\includegraphics[width=150mm,height=150mm]{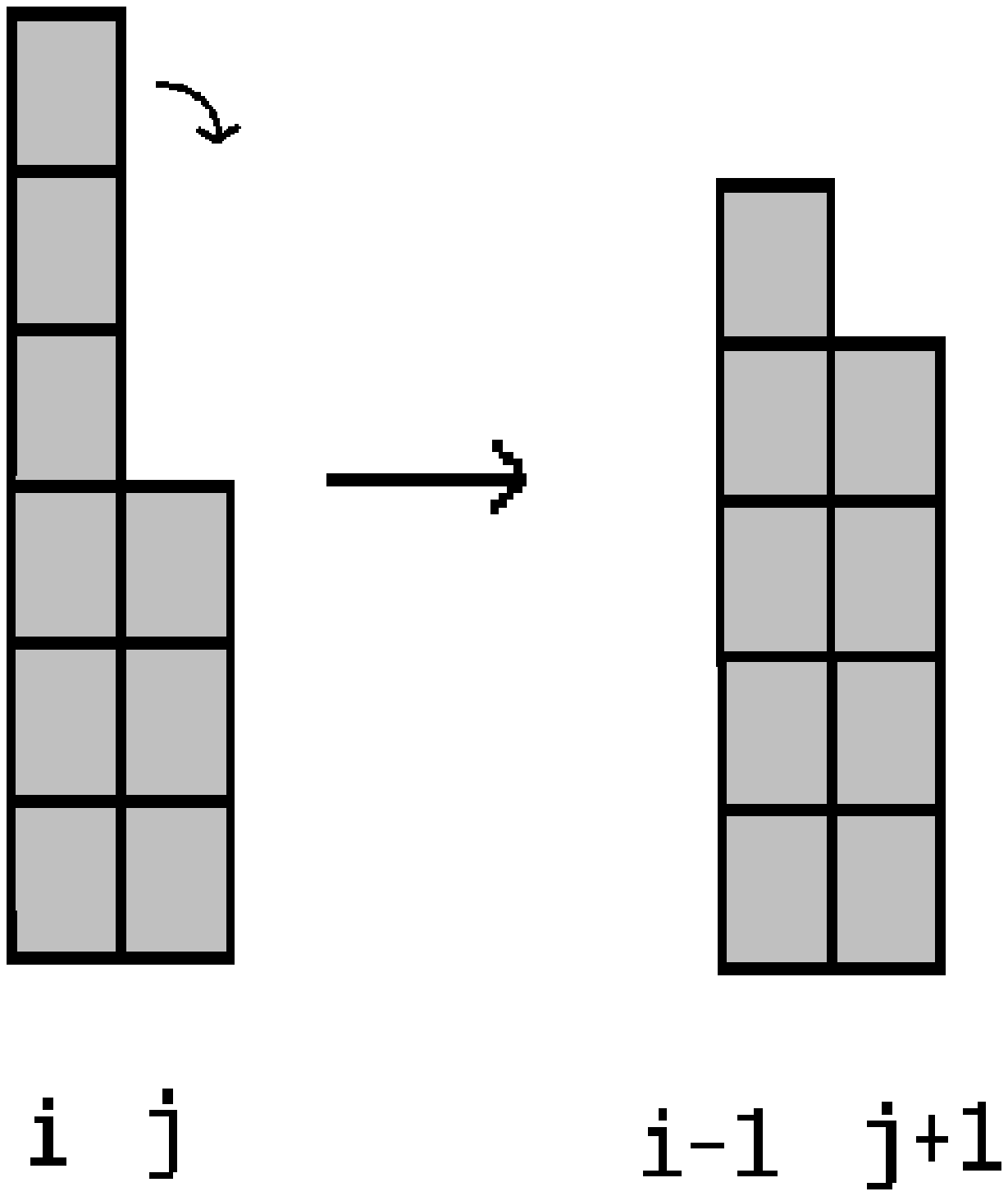}
  \caption{In the surface relaxation model,a particle drops
  with rate $\mu_{ij}$}
  \label{fig.1}
  \end{figure}
\begin{figure}
\includegraphics[width=150mm,height=150mm]{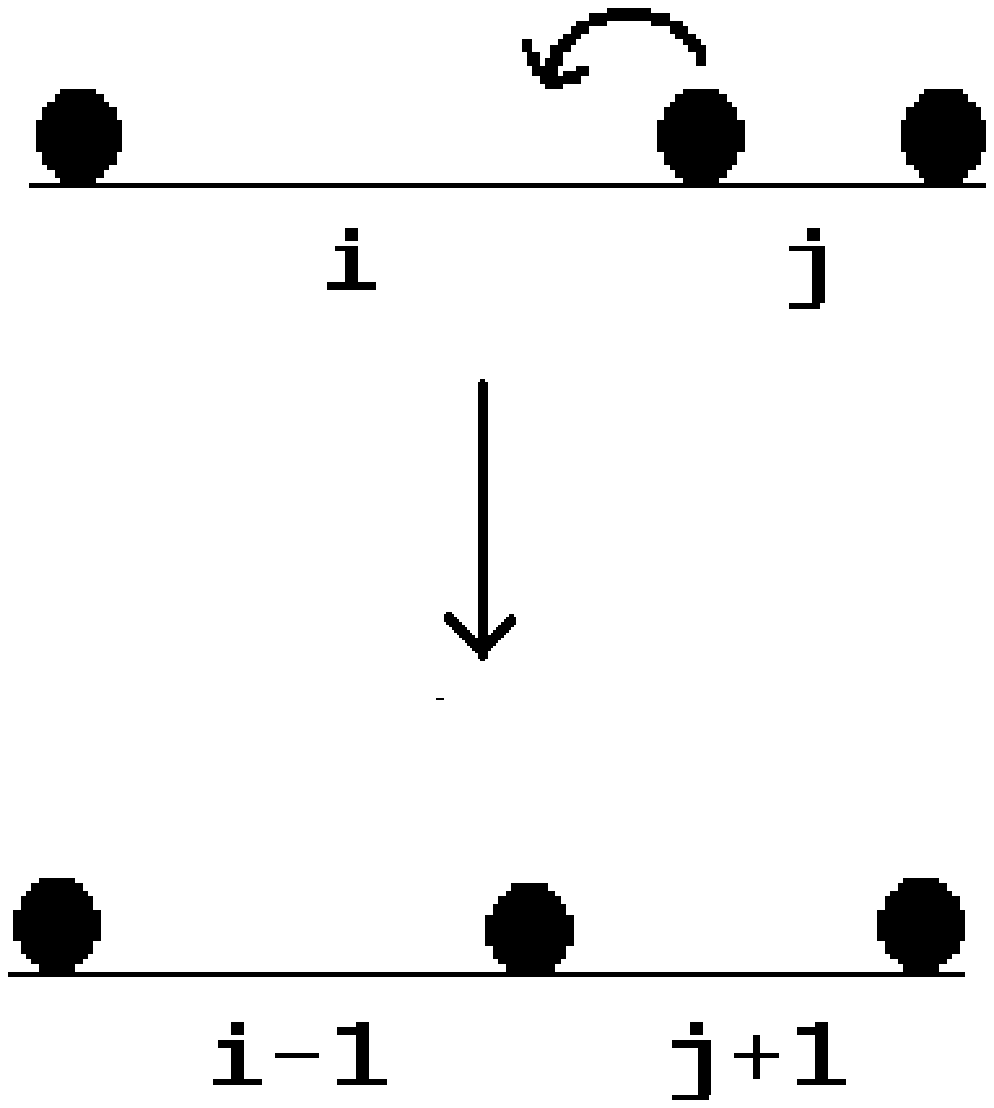}
  \caption{In the diffusive system, a particle hops with
   rate $\mu_{ij}$}\label{fig.2}
\end{figure}

\begin{figure}
\includegraphics[width=100mm,height=100mm]{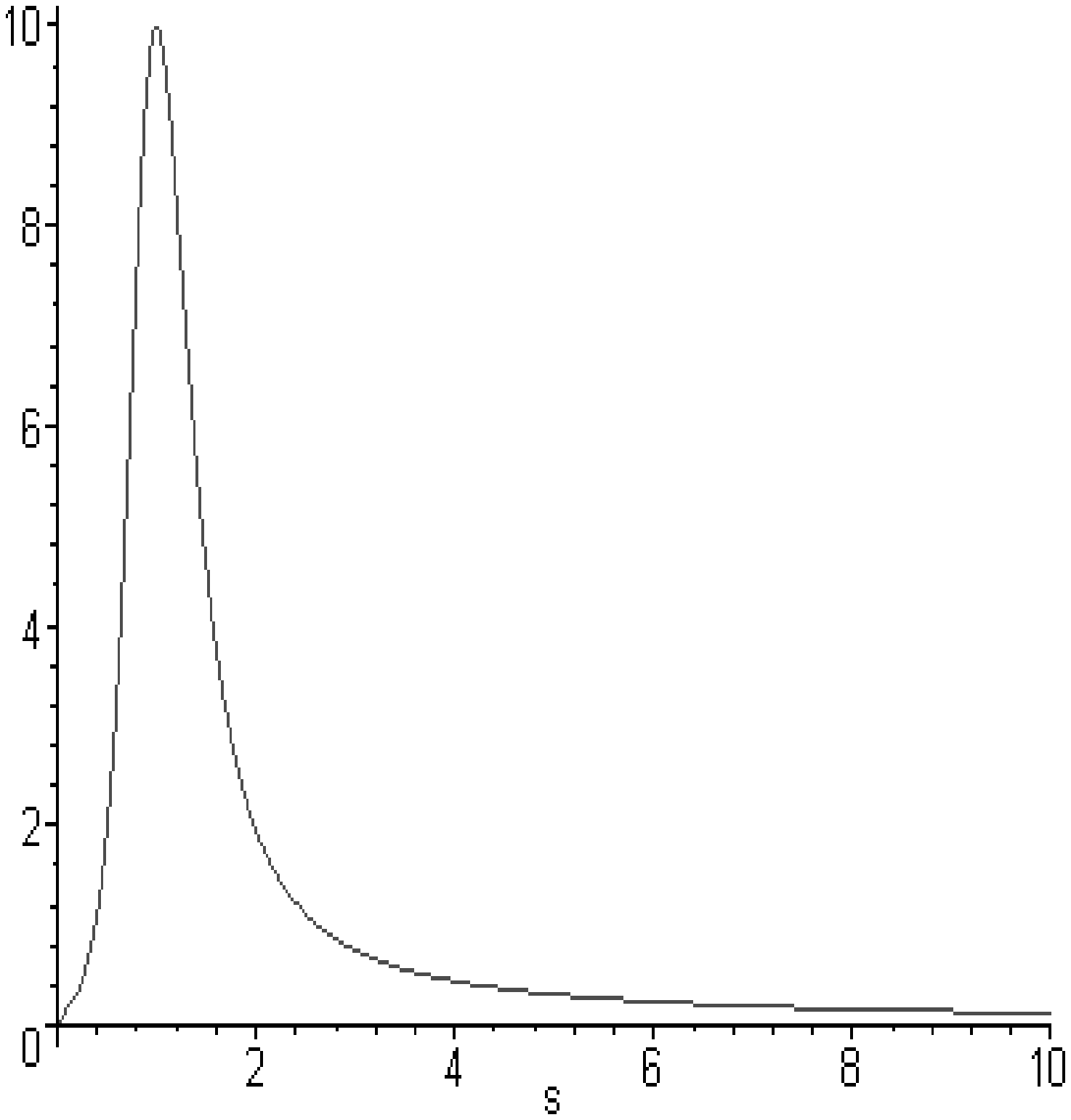}
\caption{The square of roughness for p=10} \label{fig.3}
\end{figure}

\begin{enumerate}
\bibitem{hi}H. Hinrichsen: Critical phenomena in non-equilibrium
systems, cond-mat/0001070.
\bibitem{sr}G. M. Sch\"{u}tz, in {\it Phase Transitions and
Critical Phenomena} edited by C. Domb and J. Lebowitz (Academic
Press,London, 1999).
\bibitem{sp} F. Spitzer, Adv. Math.
{\bf 5},246(1970).
\bibitem{de} B. Derrida, M.R. Evans, V.Hakim and V. Pasquier,
J. Phys. A: Math. Gen. {\bf 26},1493, (1993).
\bibitem{ahr} P. Arndt, T. Heinzel and V. Rittenberg; J. Phys. A;
Math. Gen. {\bf 31}, 833,(1998).
\bibitem{ev} M. R. Evans; J. Phys. A: Math. Gen.{\bf 30}, 5669 (1997);
Europhys. Lett.{\bf 36},13, (1996).
\bibitem {evg} M. R. Evans, D. P. Foster, C. Godreche D. Mukamel; J. Stat. Phys. {\bf
80} (1995); Phys. Rev. Lett. {\bf 74}, 208 (1995).
\bibitem{li} H.W. Lee, v. Popkov, and D. Kim, J. Phys. A {\bf
30}, 8497 (1997).
\bibitem{k1} V. Karimipour, Phys. Rev. E {\bf 59}205 (1999).
\bibitem{k2} V. Karimipour, Europhys. Letts. {\bf 47}(3), 304(1999).
\bibitem{kk} M. Khorrami, V. Karimipour; J. Stat. Phys. {\bf 100}
5/6,(2000) 999.
\bibitem{deg} J. de Gier and B. Nienhuis, Phys. Rev. E {\bf 59},4899,(1999).
\bibitem{ks} K. Krebs and S. Sandow; J.Phys. A ; Math. Gen. {\bf 30}
3165(1997).
\bibitem{adr} F. C. Alcaraz, S. Dasmahapatra and V. Rittenberg V
J.Phys. A ; Math. Gen. {\bf 31} 845 (1998).
\bibitem{r2}A. P. Isaev, P. N. Pyatov, and V. Rittenberg, Diffusion
algebras, to be published.
\bibitem{evr}M. R. Evans; Phase transitions in one-dimensional
non-equilibrium systems, cond-mat/0007293.
\bibitem{helb}D. Helbing, Traffic and related self-driven many
particle systems, cond-mat/0012229.
\bibitem{chad}D. Chowdhury, L. Santen, and A. Schadschneider,
{\it Phys. Rep.}{\bf 329},199 (2000).
\bibitem{st}A. L. Barabasi, and H. E. Stanley; {\it Fractal
Concepts in Surface Growth}Cambridge University Press, (1994).
\end{enumerate}
\end{document}